\documentclass[aps,prb,twocolumn,showpacs,psfig,superscriptaddress,longbibliography]{revtex4-1}

\usepackage{textcomp}
\usepackage{times}
\usepackage{graphicx}
\usepackage{float}
\usepackage{latexsym,amsmath,amssymb,bm,euscript}
\usepackage{color}
\usepackage{subfigure}
\usepackage{epstopdf}
\usepackage[colorlinks=true,linkcolor=blue,citecolor=blue]{hyperref}
\usepackage{hyperref}
\usepackage{soul}
\usepackage[normalem]{ulem}
\usepackage{mathrsfs}
\usepackage{amsmath}
\usepackage{lettrine}
\usepackage{xspace}
\usepackage{textcomp}

\begin{document}

\title{NMR study of supersolid phases in the triangular-lattice antiferromagnet Na$_2$BaCo(PO$_4$)$_2$}

\author{Xiaoyu~Xu}
%\thanks{These authors contributed equally to this study.}
\affiliation{School of Physics and Beijing Key Laboratory of
Opto-electronic Functional Materials $\&$ Micro-nano Devices, Renmin
University of China, Beijing, 100872, China}

\author{Zhanlong Wu}
\affiliation{School of Physics and Beijing Key Laboratory of
Opto-electronic Functional Materials $\&$ Micro-nano Devices, Renmin
University of China, Beijing, 100872, China}

\author{Ying Chen}
\affiliation{School of Physics and Beijing Key Laboratory of
Opto-electronic Functional Materials $\&$ Micro-nano Devices, Renmin
University of China, Beijing, 100872, China}

\author{Qing Huang}
\affiliation{Department of Physics and Astronomy, University of Tennessee, Knoxville, TN 37996-1200, USA}

\author{Ze Hu}
\affiliation{Institute of High Energy Physics, Chinese Academy of Sciences (CAS), Beijing 100049, China}
\affiliation{Spallation Neutron Source Science Center, Dongguan 523803, China}

\author{Xinyu Shi}
\affiliation{School of Physics and Beijing Key Laboratory of  Opto-electronic Functional Materials $\&$ Micro-nano Devices, Renmin
University of China, Beijing, 100872, China}

\author{Kefan Du}
\affiliation{School of Physics and Beijing Key Laboratory of  Opto-electronic Functional Materials $\&$ Micro-nano Devices, Renmin
University of China, Beijing, 100872, China}

\author{Shuo Li}
\affiliation{Beijing National Laboratory for Condensed Matter Physics and Institute of Physics, Chinese Academy of Sciences, Beijing, 100190, China}

\author{Rui Bian}
\affiliation{School of Physics and Beijing Key Laboratory of  Opto-electronic Functional Materials $\&$ Micro-nano Devices, Renmin
University of China, Beijing, 100872, China}

\author{Rong~Yu}
\affiliation{School of Physics and Beijing Key Laboratory of  Opto-electronic Functional Materials $\&$ Micro-nano Devices, Renmin
University of China, Beijing, 100872, China}
\affiliation{Key Laboratory of Quantum State Construction and Manipulation (Ministry of Education),
Renmin University of China, Beijing, 100872, China}

\author{Yi~Cui}
\email{cuiyi@ruc.edu.cn}
\affiliation{School of Physics and Beijing Key Laboratory of  Opto-electronic Functional Materials $\&$ Micro-nano Devices, Renmin
University of China, Beijing, 100872, China}
\affiliation{Key Laboratory of Quantum State Construction and Manipulation (Ministry of Education),
Renmin University of China, Beijing, 100872, China}

\author{Haidong~Zhou}
\email{hzhou10@utk.edu}
\affiliation{Department of Physics and Astronomy, University of Tennessee, Knoxville, TN 37996-1200, USA}

\author{Weiqiang~Yu}
\email{wqyu\_phy@ruc.edu.cn}
\affiliation{School of Physics and Beijing Key Laboratory of  Opto-electronic Functional Materials $\&$ Micro-nano Devices, Renmin
University of China, Beijing, 100872, China}
\affiliation{Key Laboratory of Quantum State Construction and Manipulation (Ministry of Education),
Renmin University of China, Beijing, 100872, China}

%\date{\today}

\begin{abstract}
We report ultra-low-temperature $^{23}$Na NMR measurements on the Ising triangular lattice antiferromagnet Na$_2$BaCo(PO$_4$)$_2$,
which precisely resolve the phase diagram under magnetic field applied along the crystalline $c$ axis.
With increasing field, the NMR spectra
resolve three ordered phases with distinct spin configurations: the Y, up-up-down (UUD), and V phases.
The spin-lattice relaxation rate $1/T_1$ data demonstrate gapless excitations in the Y and V phases,
strongly supporting their supersolid nature.
However, the phase transitions from the UUD phase to the two supersolid phases exhibit dramatically different behaviors upon cooling.
Prior to entering the Y phase,
$1/T_1$ identifies a gapless regime within the UUD phase, suggesting a
Berezinskii-Kosterlitz-Thouless phase above a second-order phase transition.
In contrast, the coexistence of the UUD and V phases observed in our experiments provides direct evidence of a first-order phase transition between these phases.
\end{abstract}

\maketitle

{\it Introduction.---} Supersolid is a state of matter that exhibits both the incompressibility of a solid and the frictionless flow of a superfluid.
It was originally proposed in the $^{4}$He system~\cite{andreev_JETP_1969,Thouless_AoP_1969,Leggett_PRL_1970}, where vacancies~\cite{Nikolay_AiP_2007},
disorder~\cite{Boninsegni_PRL_2007,Pollet_PRL_2007}, or the pinning by $^{3}$He~\cite{Day_Nature_2007,Day_PRB_2009,Balibar_Nature_2010}
may account for the realization of supersolidity.
However, experimentally, the existence of supersolid in this system remains highly debated~\cite{Kim_Nature_2004,Kim_PRL_2012}.
In cold quantum gases, the supersolid phase was also theoretically proposed and described by the
hard-core boson model~\cite{Baym_PRA_2012,Pitaevskii_Pramana_1987,Ancilotto_PRB_2005,Norcia_Nature_2021}.
Notably, the coexistence of solid and superfluid order was observed in a Bose-Einstein condensates of
cold atoms with spin-orbit coupling, cooled in optical lattices~\cite{Li_Nature_2017, Recati_NRP_2023}.
By mapping bosons to spin-1/2 magnetic systems~\cite{Massimo_PRL_2005,Wang_PRL_2009}, a spin supersolid phase~\cite{Paramekanti_PRL_2010} is also proposed
to emerge in frustrated magnets~\cite{Wessel_PRL_2005}. For example, in an Ising triangular-lattice antiferromagnet (TLAFM) with the field applied
along the magnetic easy axis, the spin supersolid phase emerges through two simultaneous symmetry-breaking processes: the breaking of in-plane spin rotational symmetry via Bose-Einstein condensation to form a spin superfluid, and the breaking of out-of-plane translational symmetry to establish a spin solid order~\cite{Wang_PRL_2009}.

Recently, neutron scattering and thermodynamic studies on several Ising cobaltate TLAFMs,
such as Na$_2$BaCo(PO$_4$)$_2$~\cite{Zhang_PRL_2024, xiang_Nature_2024}
and K$_2$Co(SeO$_3$)$_2$~\cite{Zhong_PRM_2020, chen_arXiv_2024}, have observed field-induced coexistence
of $\sqrt{3}$$\times$$\sqrt{3}$ magnetic sublattices, representing solid order, and gapless Goldstone excitations,
representing superfluid order. These findings provide compelling evidence for the existence of spin supersolid
phase~\cite{Zhong_PRM_2020, Zhang_PRL_2024, xiang_Nature_2024, chen_arXiv_2024}.
For Na$_2$BaCo(PO$_4$)$_2$~\cite{Li_NC_2020,Wellm_PRB_2021,Sheng_2022_PNAS}, which is described by a spin-1/2 XXZ model.
The system hosts two field-induced spin supersolid phases:
the Y phase at low fields and the V phase at high fields~\cite{Gao_npj_2022}.
Here, weak easy-axis anisotropy enables a large regime of supersolid phases alongside the up-up-down (UUD) phase~\cite{Yamamoto_PRL_2014}.
The giant magnetocaloric effect critical point between the V phase and the fully polarized (FP) phase offers
the potential as an ultra-low-temperature coolant~\cite{xiang_Nature_2024,Zhang_PRL_2024},
harnessing spin fluctuations over a broad field range due to magnetic frustration.
Recent tensor network calculations have identified roton-like excitations in the Y phase,
which are proposed to originate from a magnon-Higgs mechanism.
However, these features remain unconfirmed in the V phase~\cite{chi_arxiv_2024}.

\begin{figure}[t]
\includegraphics[width=8.5cm]{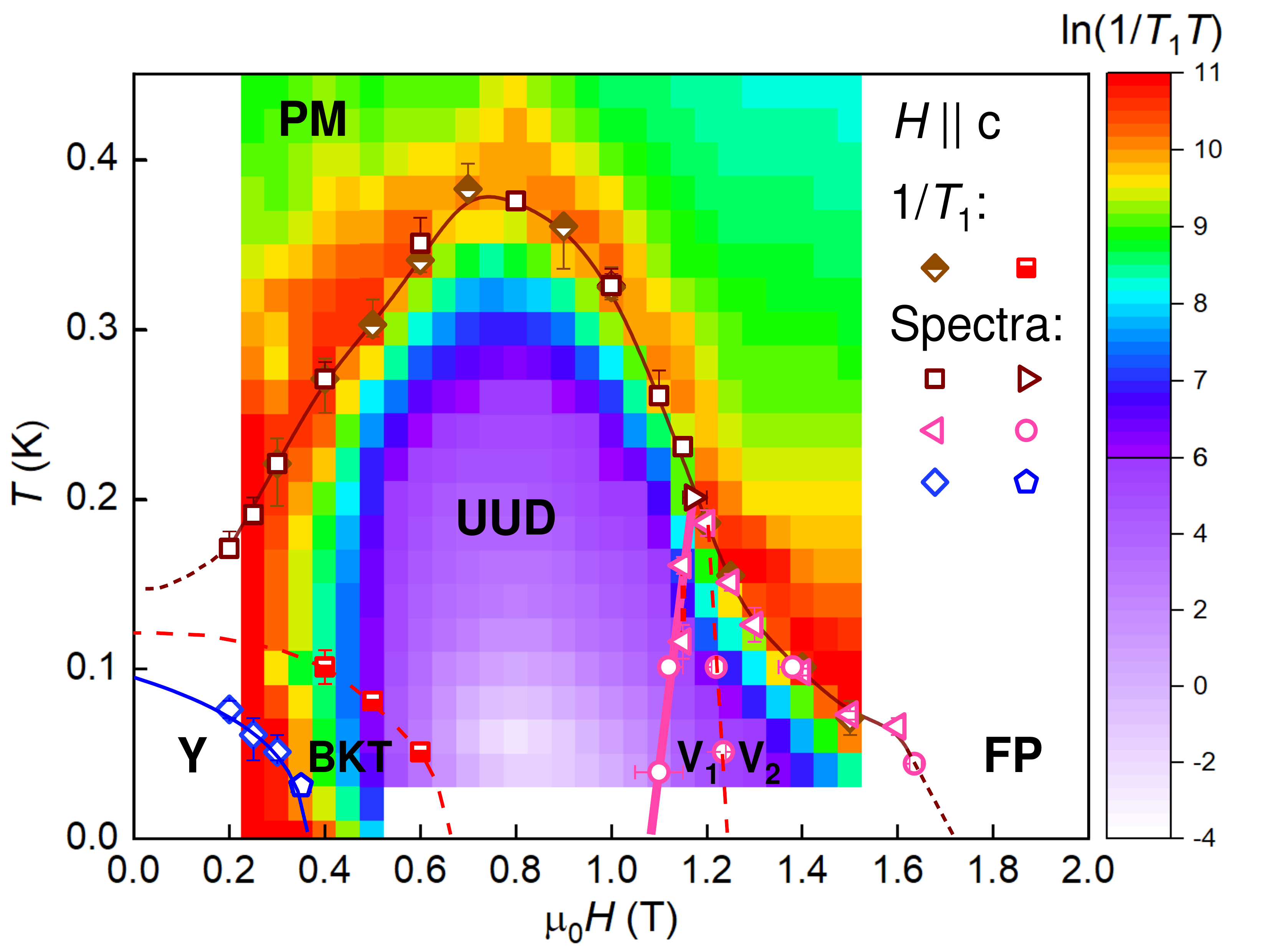}
\caption{\label{pd}Magnetic phase diagram.
The color map depicts a contour plot of the $1/T_1T$ data. The PM, UUD, Y, V,
and FP phases and the phase transition lines among them are also determined.
V$_{\rm 1}$ and V$_{\rm 2}$ are V phases with different configurations.
Phase boundaries were determined using different methods
as indicated. Thick (thin) solid lines represent the first (second)-order
phase transitions, while dashed lines above the BKT phase
and on the right side of the V$_{\rm 1}$ phase represent phase crossover lines
with no obvious phase transition.
}
\end{figure}

In this study, we conducted ultra-low-temperature NMR measurements on Na$_2$BaCo(PO$_4$)$_2$, with the field applied along
the $c$ axis. By analyzing NMR spectra at the lowest temperature,
we precisely mapped the phase boundaries of the Y, UUD, and V phases, as shown in Fig.~\ref{pd}.
Gapless excitations were identified in the Y and V phases,
confirming their supersolid nature, adjacent to the gapped UUD phase. More interestingly, a
large gapless region above the UUD-Y phase boundary was observed, indicating the formation of a two-dimensional (2D) Berezinskii-Kosterlitz-Thouless (BKT)-like phase.
In contrast to the second-order UUD-Y phase transition found in our experiments,
the UUD-V transition is identified as a first-order transition, evidenced by the coexistence of the UUD and V phases.
Enhanced spin fluctuations were also revealed near the phase boundaries of the Y and V phases,
as demonstrated by the color map of $1/T_1T$ in the phase diagram.
Our identification of the BKT phase, the first-order transition boundary,
and the strong spin fluctuations offers new insights into the supersolid phase in this system.

High-quality single crystals were grown by the flux method~\cite{Zhong_PNAS_2019}.
The Co$^{2+}$ ions, enclosed in the CoO$_6$ octahedra, form layered spin-1/2 equilateral
triangular lattices with an AA-stacked structure along the $c$ axis.
Na$^{+}$ ions are positioned at interstitial sites within the interstitial planes, located directly above or below the centers of the Co$^{2+}$ triangles.
We performed $^{23}$Na ($I$=3/2, Zeeman factor $\gamma$=11.262~MHz/T) NMR measurements in a dilution refrigerator
with temperature down to 0.024~K. The NMR spectra were collected using the standard spin-echo technique. The spin-lattice relaxation time
$T_1$ was obtained using the inversion-recovery method.

\begin{figure}[t]
\includegraphics[width = 8.8cm]{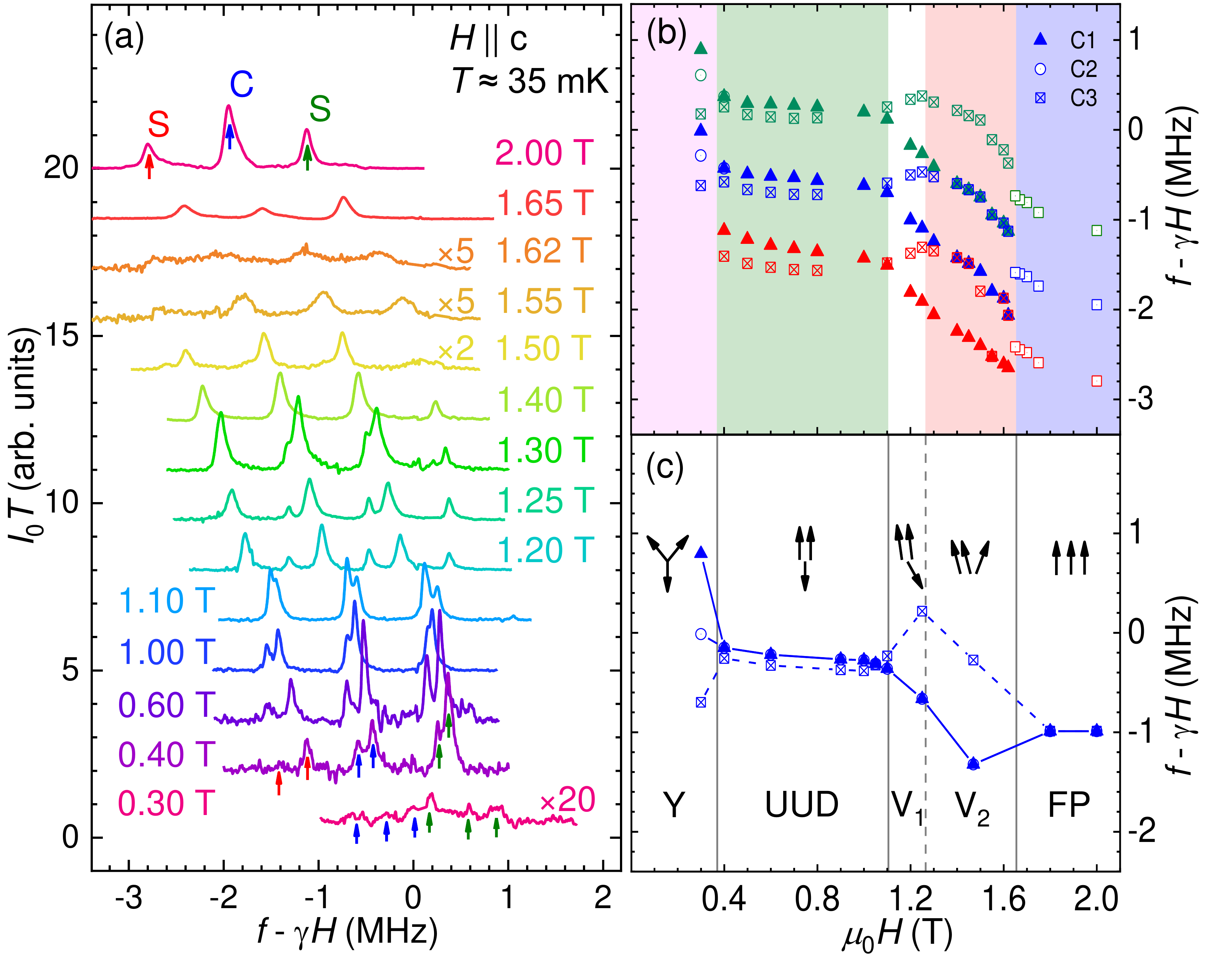}
\caption{\label{f50mK}$^{23}$Na Spectra at 0.035~K.
(a) Spectra taken with decreasing fields. Data are shifted vertically for clarity.
Near the transition fields, spectra are scaled by factors as marked.
C and S label the central and satellite peaks, respectively.
(b) Peak frequencies as functions of fields, where blue, green and red symbols represent
the central (C) and satellite peaks (S), respectively, as labeled in (a).
C1, C2 and C3 correspond to three central lines split due to magnetic ordering.
(c) The resonance frequencies of simulated central NMR lines in the Y, UUD, V$_1$, V$_2$, and FP phases (see text) are plotted, with their respective magnetic configurations in each Co$^{2+}$ triangle, as illustrated by the arrows.
Vertical lines mark the phase boundaries between different phases.
}
\end{figure}

{\it NMR Spectra.---} To resolve the distinct field-induced phases,
the spectra were collected at temperatures around 0.035~K under varying magnetic fields, as depicted in Fig.~\ref{f50mK}(a).
The spectrum at 2~T exhibits one central peak (labeled as C) and two satellite peaks (labeled as S)
as expected for the FP phase above the critical field,
consistent with uniform magnetization.
The separation between each satellite and the central line is approximately $\nu_{\rm}\approx$~0.85~MHz.
At 1.62~T, the spectral weight is significantly reduce, indicating the onset of the
critical field, consistent with prior report~\cite{Zhong_PNAS_2019}.
At lower field, more NMR lines emerge,
indicating the onset of magnetic ordering through NMR line splitting.
Another quantum phase transition occurs near 0.3~T, where the spectral weight
is significantly reduced again.

\begin{figure}[t]
	\includegraphics[width=8.5cm]{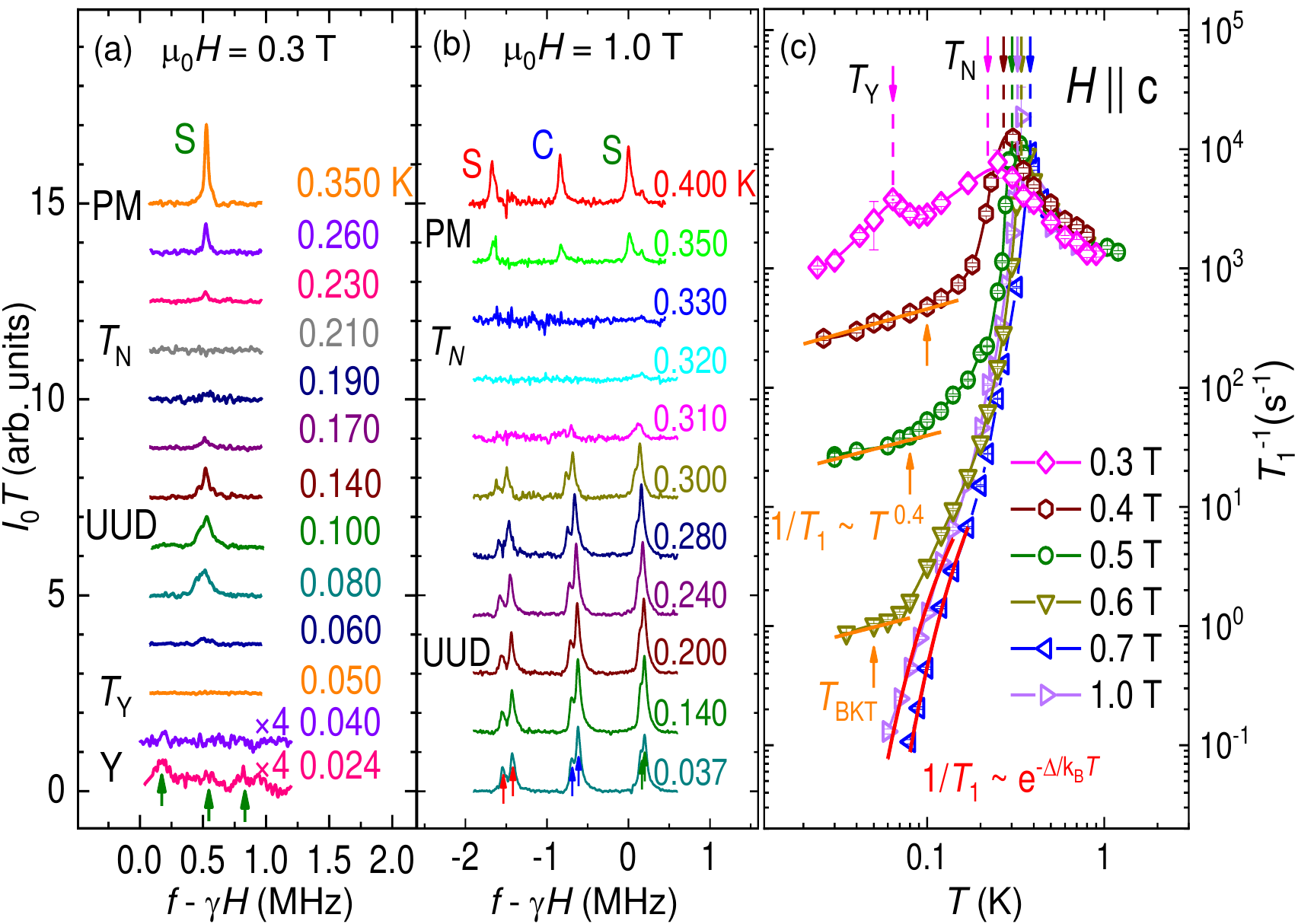}
	\caption{\label{yuud}NMR spectra at low fields.
 (a) High-frequency satellite spectra measured over a range of temperatures at 0.3~T, with the PM, UUD and Y phases and phase transition temperatures as labeled.
 (b) Full spectra measured at typical temperatures at 1~T, exhibiting a single PM-UUD phase transition.
 (c) $1/T_1$  as functions of temperatures at typical fields. $T_{\rm N}$ and $T_{\rm Y}$ denote
 the phase transitions into the UUD phase and Y phase, respectively.
 The gap function fits in the UUD phase are represented by the red solid lines, and the power-law function fits are depicted by straight lines.
}
\end{figure}

To resolve different magnetic phases under an applied field,
the resonance frequencies of all peaks are plotted as functions of
fields as shown in Fig.~\ref{f50mK}(b).
We also simulated the NMR spectra for the Y, UUD, V$_1$, V$_2$ and FP phases,
with the spin configurations illustrated in Fig.~\ref{f50mK}(c).
These simulations assume dipolar-type hyperfine coupling between $^{23}$Na and Co$^{2+}$ moments.
The resonance frequencies of the simulated central peak of $^{23}$Na are shown in Fig.~\ref{f50mK}(c),
where a 3$^\circ$ field deviation from the $c$ axis is applied to account for the observed spectra.
By comparison, our data are in good agreement with these field-induced phase with increasing field,
where three central peaks are observed in the Y phase (labeled as C1, C2 and C3),
two in the UUD and the V phase (labeled as C1 and C3), and a single
peak in the FP phase. Additionally, abrupt changes in the slope of the resonance frequency are observed at the Y-UUD transition (around 0.37 T), the UUD-V transition (around 1.12 T), and the V-FP transition (around 1.65 T).

Here we also found that, although both configurations belong to the same
V phase, a crossover from the V$_1$ configuration to the V$_2$ configuration can be distinguished,
characterized by different canting of the solid order and a non-monotonic change of the resonance
frequency of C3, as shown in Fig.~\ref{f50mK}(b) and (c).
Based on this, we measured the spectra at several constant temperatures
and determined the left and right boundaries of the V$_1$ phase,
as shown in the phase diagram in Fig.~\ref{pd}.

{\it Y--UUD phase transition.---} In this compound, the transition temperatures can be determined by both the
NMR spectra and the spin-lattice relaxation rate.
To resolve the phase transitions at low fields, we measured the high-frequency satellite spectra, whose
frequency range is accessible within the operating window of our NMR spectrometer (above 3.5~MHz).
At 0.3~T, as shown in Fig.~\ref{yuud}(a), the high-frequency satellite spectra
reveal two phase transitions upon cooling, at $T_{\rm N}\approx$~0.21~K
and $T_{\rm Y}\approx$~0.05~K, respectively, consistent with previous reports~\cite{xiang_Nature_2024,Sheng_2022_PNAS}.
A significant reduction in spectral weight is observed at each temperature,
corresponding to the paramagnetic (PM)-to-UUD phase transition and the UUD-to-Y phase transition respectively.
The signal is largely wiped out due to strong low-energy spin fluctuations at these critical temperatures.
In the Y phase, the satellite also exhibit three NMR peaks, as seen in the central NMR spectra in Fig.~\ref{f50mK}(a).

At 1~T, as illustrated in Fig.~\ref{yuud}(b),
the system enters the UUD phase from the PM phase at approximately 0.325~K~(marked by $T_{\rm N}$),
where six distinct peaks are resolved.
The spectrum splits into two sets of peaks comprising two central lines and four satellite lines.
The two central lines, as marked by the blue arrows with a notably small frequency split,
exhibit a spectral weight ratio of 2:1, consistent with expectations for the UUD phase.
Specifically, two up moments induce a higher resonance frequency at their nearest Na sites.

The low-energy spin fluctuations are studied using the $1/T_1$.
The $1/T_1$ is measured on the high-frequency satellite peak at low fields (below 0.4~T) and on the central peak at high fields.
Figure~\ref{yuud}(c) demonstrates typical $1/T_1$ data as functions of temperatures with fields ranging from 0.3~T to 1~T.

At 0.3~T, the $1/T_1$ exhibits two peaks as the temperature decreases,
indicating that the system undergoes two magnetic phase transitions, as seen in the spectra.
The first peak at 0.25~K, marked by $T_{\rm N}$, signals the PM-to-UUD phase transition.
By this,  $T_{\rm N}$ is determined in all fields and is shown in the phase diagram in Fig.~\ref{pd}.
A second peak, observed at 0.065~K as marked by $T_{\rm Y}$ at 0.3~T, is the onset of the phase transition from the
UUD to the Y phase. The slow decrease of $1/T_1$ from $T_{\rm Y}$  down to 0.024~K indicates that the system enters
a gapless phase, strongly supporting the presence of the supersolid phase with spontaneous breaking of U(1) symmetry.

At 0.7~T and 1 T, a gap opening behavior below $T_{\rm N}$ is clearly observed, as expected for the
magnetization plateau phase, that is, the UUD phase.
As demonstrated in Fig.~\ref{yuud}(c), the data are fit to the function $1/T_1$$\sim$$e^{-\Delta/k_BT}$,
where $\Delta$ is the spin gap. From the fit, the gap is obtained as 0.66$\pm$0.01~K and
0.44$\pm$0.01~K at 0.7~T and 1.0~T respectively.

At 0.4~T and 0.6~T, a peaked behavior is still observed at $T_{\rm N}$, marking the transition to the UUD phase.
Below $T_{\rm N}$, $1/T_1$ initially drops rapidly. However, far below $T_{\rm N}$, a level-off behavior is
observed, despite the absence of Y-phase signatures in the spectra at low temperatures.
In fact, the low-temperature data follow a power-law behavior, $1/T_1{\sim}T^{0.4}$,
as demonstrated by the fits in Fig.~\ref{yuud}(c). We then determine the onset temperature of the power-law behavior at
different fields, as marked in Fig.~\ref{yuud}(c). The onset temperature is about 0.1~K at 0.4~T
and decrease to 0.05~K at 0.6~T. These temperatures are plotted in the phase diagram as shown in Fig.~\ref{pd},
 which are positioned right above the Y phase in field and temperature.

Such a power-law behavior of $1/T_1$ indicates a gapless state, contrasting with the gapped behavior expected
for the UUD phase. Because this behavior occurs at fields right above the Y phase, we believe it evidences the existence
of a quasi-long range order of the Y phase in the system. As no signature of the 3D Y-type order is observed,
this aligns with the presence of a BKT phase, which maintains the U(1) symmetry within the $ab$ plane.
Further reduction of temperature and magnetic field leads to the emergence of the Y phase, accompanied by the spontaneous breaking of U(1) symmetry, as depicted in the phase diagram.

\begin{figure}[t]
\includegraphics[width=8.5cm]{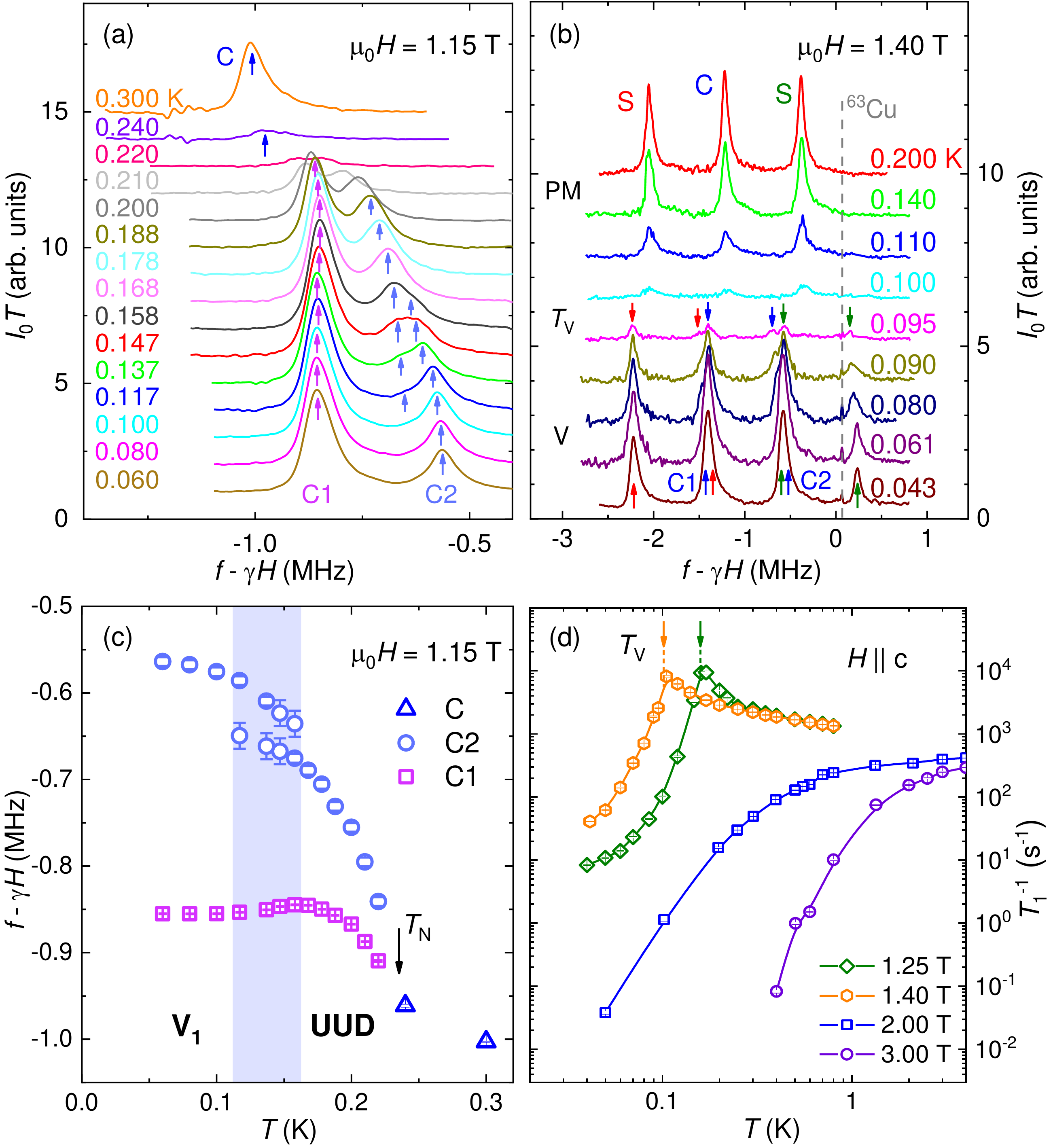}
\caption{\label{uudv}Spectra at high fields.
(a) Spectra of the central peak measured at 1.15~T. The peak positions are marked by arrows.
(b) Full spectra measured at 1.4~T. The blue arrows mark the central peaks, and the red and green arrows mark the satellite peaks.
(c) Resonance frequencies of peaks extracted from (a), plotted as functions of field. The shaded blue areas mark the range
of phase coexistence.
(d) $1/T_1$ data at typical fields in the V phase and the FP phase.
}
\end{figure}

{\it UUD--V phase transition.---} To resolve the nature of the phase transition between
the UUD and the V phase, we performed spectral
measurements at 1.15~T upon cooling (Fig.~\ref{uudv}(a)),
and extracted the peak frequencies, which are plotted in Fig.~\ref{uudv}(c).
The N\'{e}el transition at $T_{\rm N}$ is again determined at about 0.22~K, as indicated by spectral intensity loss.
While a double-peak feature is observed in the spectra at temperatures above 0.158 K and below 0.1 K, a three-peak feature emerges at intermediate temperatures between 0.158 K and 0.1 K.

Note that the spectra of both the UUD phase and the V phases, as shown in Fig.~\ref{f50mK}(b),
split into two sets of peaks. To further confirm this, spectra at a higher field of 1.4~T
were measured upon cooling and are shown in Fig.~\ref{uudv}(b). The N\'{e}el transition is observed
at about 0.1~K, marked by a significant loss in spectral intensity. At 0.095~K and below,
all peaked spectra split into two sets, as indicated by the blue arrows, suggesting the
presence of a single V phase rather than a two step phase transition in this field.
To confirm this, the $1/T_1$ data are shown in Fig.~\ref{uudv}(d), which resolves
the field regime of the supersolid phase. By comparing the $1/T_1$ data at different fields,
gapped behavior is observed at 1.0~T for the UUD phase and at 2.0~T for the FP
phase. At 1.25~T and 1.40~T, a gapless behavior is seen at
the lowest temperatures, confirming the gapless nature of the V phase.

As 1.15~T is close to the transition region between
the UUD and V phases, we believe the three-peak feature observed
at intermediate temperatures, as shown in Fig.~\ref{uudv}(a),
is caused by the phase separation of the UUD and the V phase
at the phase boundary. Such phase separation
indicates a first-order phase transition from the UUD phase to the V phase
upon cooling.

{\it Phase diagram and discussions.---}
With all phase boundaries determined, a full ($H$, $T$) phase diagram is constructed as
shown in Fig.~\ref{pd}. A colored contour plot of ${\rm ln}(1/T_1T)$ data is also shown for fields ranging
from 0.25~T to 1.5~T. Our data resolve four transition lines (solid lines), including the PM--UUD, UUD--Y,
PM--V, and UUD-V transitions. In addition to the algebraic BKT phase
reported at zero field~\cite{xiang_Nature_2024}, the existence of another BKT phase at finite
fields has been added to the phase diagram.
The false-color map of $1/T_1T$ also demonstrates spin fluctuations at the transitions
between PM--UUD, PM--V, and UUD--Y.

In particular, the color map highlights a broad region of low-energy spin fluctuations at temperatures
just above the Y and V phase.
For instance, the extent of the temperature range with strong fluctuations
above the upper boundaries of both the Y and V phases, such as at 0.25~T and 1.5~T, spans
at least 50\% of the corresponding $T_{\rm Y}$ and $T_{\rm V}$. This further demonstrates that the supersolid phase
could be used for ultra-low-temperature magnetocaloric cooling over a large temperature range~\cite{xiang_Nature_2024}.
These effects emphasize that magnetic frustration enlarges the regime of magnetic
fluctuations due to phase competition or ground state degeneracy.

The phase diagram also reveals an asymmetry in the phase boundaries of the two supersolid phases. Specifically, a BKT-like phase transition is observed in the UUD phase,
which lies directly above the Y phase, whereas no such transition is observed in the V phase.
The UUD-Y phase transition is characterized by second-order behavior,
whereas the UUD-V transition follows a first-order phase transition.
In contrast, theories based on the 2D model predicts both transitions to be of the second-order
type~\cite{Yamamoto_PRL_2014,Gao_npj_2022}. These discrepancies suggest the need for additional magnetic interactions to account for our experimental observations.
For instance, interlayer coupling or Dzyaloshinskii-Moriya (DM) interactions
may introduce stronger phase competition between the UUD and V phases,
thereby driving a first-order transition.

Indeed, our data suggest that interlayer couplings play different roles
in different field regimes. At low field, the effective interlayer
coupling appears very weak, as evidenced by the large regime
of BKT phase above the Y phase, consistent with previous reports of weak interlayer
couplings~\cite{Zhong_PNAS_2019,Sheng_2022_PNAS,Gao_npj_2022}. Conversely,
the observation of a first-order-like phase transition between the UUD and V phases
may suggests that interlayer coupling becomes significant at high fields.
Recent numerical calculations on the triangular lattice antiferromagnet
reveal that antiferromagnetic interlayer coupling can lead to first-order
phase transitions near a high-field supersolid phase, even though the system has an easy-plane anisotropy~\cite{Yamamoto_2025_PRL}, rather than the easy-axis anisotropy
in the current case.

Note that the BKT phase has also recently been reported in several materials,
such as the Ising TLAFM TmMgGaO$_4$~\cite{Hu_2020_NC} and
the XY honeycomb-lattice antiferromagnet NiPS$_3$~\cite{Seifert_2022_PRB,LC_2024_PRB}. These
findings suggest that magnetic frustration may significantly suppress the magnetic phase
transition temperatures, thereby enhancing the likelihood of experimental observation.
We also note that a recent NMR study on Na$_3$Ni$_2$BiO$_6$, a frustrated honeycomb-lattice antiferromagnet
with easy-axis anisotropy, reported strong low-energy spin fluctuations in the
field-induced magnetic phases~\cite{Shi_2024_CPB}, which may also suggest the
existence of a BKT phase near a supersolid state.

{\it Summary.---} We investigated the supersolid phase in high-quality single crystals
of the Ising TLAFM Na$_2$BaCo(PO$_4$)$_2$ and precisely constructed its phase diagram under field.
Gapless excitations were observed in the Y and V phases, providing strong evidence of their supersolid nature.
Strong low-energy spin fluctuations persist at temperatures up to 50\% above the critical transitions of both supersolid phases. Asymmetric behavior is also observed in the transitions
from the UUD phase to both supersolid phases. The phase transition between the UUD and Y
phases is of second-order type, accompanied by a nearby BKT phase.
In contrast, the phase transition between the UUD and V phases exhibits first-order behavior,
suggesting that additional interactions, particularly interlayer coupling, may be effective in
enhancing spin fluctuations and altering the nature of phase transitions.

{\it Acknowledgements.---} This work is supported by the National Key Research and
Development Program of China (Grant No. 2023YFA1406500), and the National Natural Science
Foundation of China (Grants Nos.~12374156, 12134020, and 12174441).
Q.H. and H.D.Z. thank the support from the National Science Foundation Grant No. DMR-2003117.

\end{document}